\documentclass[a4paper,fleqn,usenatbib]{mnras}

\usepackage{newtxtext,newtxmath}

\usepackage[T1]{fontenc}
\usepackage{ae,aecompl}

 \usepackage{amssymb,amsmath,fancyhdr,graphicx,lscape,multirow,deluxetable}

\title[The Invisible AGN Catalogue]{The Invisible AGN Catalogue: A Mid-Infrared -- Radio Selection Method for Optically-Faint Active Galactic Nuclei}

\author[Truebenbach \& Darling.]{
Alexandra E. Truebenbach,$^{1}$\thanks{E-mail: alexandra.truebenbach@colorado.edu}
Jeremy Darling$^{1}$
\\
$^{1}$Center for Astrophysics and Space Astronomy, Department of Astrophysical and Planetary Sciences, University of Colorado, 389 UCB, Boulder, CO 80309-0389, USA
}

\date{Accepted XXX. Received YYY; in original form ZZZ}

\pubyear{2016}

\begin{document}
\label{firstpage}
\pagerange{\pageref{firstpage}--\pageref{lastpage}}
\maketitle

\begin{abstract}
A large fraction of active galactic nuclei (AGN) are ``invisible'' in extant optical surveys due to either distance or dust-obscuration. The existence of this large population of dust-obscured, infrared-bright AGN is predicted by models of galaxy -- supermassive black hole coevolution and is required to explain the observed X-ray and infrared backgrounds. Recently, infrared colour-cuts with WISE have identified a portion of this missing population. However, as the host galaxy brightness relative to that of the AGN increases, it becomes increasingly difficult to differentiate between infrared emission originating from the AGN and from its host galaxy. As a solution, we have developed a new method to select obscured AGN using their 20 cm continuum emission to identify the objects as AGN. We created the resulting Invisible AGN Catalogue by selecting objects that are detected in AllWISE (mid-IR) and FIRST (20 cm), but are not detected in SDSS (optical) or 2MASS (near-IR), producing a final catalogue of 46,258 objects. 30 per cent of the objects are selected by existing selection methods, while the remaining 70 per cent represent a potential previously-unidentified population of candidate AGN that are missed by mid-infrared colour cuts. 
Additionally, by relying on a radio continuum detection, this technique is efficient at detecting radio-loud AGN at $z \ge 0.29$, regardless of their level of dust obscuration or their host galaxy's relative brightness.

\end{abstract}




\section{Introduction}\label{intro}
Models predict that obscured (Type 2) active galactic nuclei (AGN) outnumber unobscured (Type 1) AGN by a factor of $\sim$ 3 \citep[e.g.][]{Comastrietal1995,Treisteretal2004, Ballantyneetal2011}, yet none of the current AGN detection techniques have observationally reproduced this fraction \citep[e.g.][]{Treisteretal2009b,Maliziaetal2009,Rovilosetal2014,BrandtAlexander2015}. This is primarily because the preferred selection methods rely on the detection of several optical emission lines (e.g. [OIII], [NII], H $\alpha$, and H $\beta$; \citealt{Baldwinetal1981}), thereby excluding all but the optically-brightest, most dust-poor systems \citep[e.g.][]{Reyesetal2008}.  
A new AGN detection method that is unbiased towards Type 2 AGN is needed to confirm the theoretically predicted obscured AGN fraction. This fraction is crucial for testing theories of galaxy -- supermassive black hole (SMBH) coevolution and for detecting the sources of the observed infrared and hard X-ray backgrounds \citep{FabianIwasawa1999, CharyElbaz2001, Worsleyetal2004, Worsleyetal2005, Treisteretal2009b, Shietal2013a, Shietal2013b,Rovilosetal2014}. 

The mid-infrared (mid-IR) provides a useful alternative wavelength regime that is less affected by extinction. Several AGN detection techniques have been developed with the Infrared Array Camera \citep{Fazioetal2004} on {\it Spitzer} \citep[e.g.][]{Lacyetal2004, Sternetal2005, Donleyetal2008, Donleyetal2012}. Now, the {\it Wide-field Infrared Survey Explorer} \citep[WISE;][]{Wrightetal2010} provides a high sensitivity, all-sky survey with which to use these techniques.  By combining this mid-IR survey with other wavelength regimes to remove contamination from dusty star-forming galaxies (see below), we can identify many more
obscured and unobscured AGN without a reliance on optical spectra. Since
its launch in 2009, many studies have used WISE, which is inherently
biased towards infrared-bright, dusty objects, to identify large samples of
Type 1 and Type 2 AGN \citep[e.g.][]{DAbruscoetal2012, EdelsonMalkan2012,
  Eisenhardtetal2012, Massaroetal2012, Tsaietal2014, Assefetal2015,
  Jonesetal2015, Mingoetal2016}. These identification techniques are often simple
 mid-infrared colour cuts, such as W1 $-$ W2 $\ge 0.8$ (i.e. [3.4 $\mu$m] $-$ [4.6 $\mu$m] $\ge 0.8$, where both are in Vega magnitudes). This particular colour cut, anticipated prior to the launch of WISE by
 \cite{Ashbyetal2009}, \cite{Assefetal2010}, and \cite{Eckartetal2010}, and
 realized by \cite{Sternetal2012}, is preferred because it relies only on
 the two most sensitive WISE bands.  To further reduce the contamination from star-forming galaxies, \cite{Mateosetal2012} proposed a colour cut using the first three WISE bands (3.4, 4.6, and 12 $\mu$m). \cite{Mateosetal2012} used the Bright Ultrahard {\it XMM-Newton} Survey to inform their selection criteria and found that a wedge in W1 $-$ W2, W2 $-$ W3 colour space suffers less contamination from star-forming galaxies, while only slightly reducing completeness.

The biggest challenge of identifying AGN based on their infrared emission
is distinguishing dust heated by the AGN from star formation within
the host galaxy. Although the W1 $-$ W2 colour cut is successful at identifying pure AGN, even those obscured with significant levels of dust extinction, out to at least $z \sim 3.5$, it fails to select many AGN with moderate amounts of host galaxy contamination. 
For AGN with no dust extinction, the W1 $-$ W2 colour cut begins to miss
AGN where $> 50$ per cent of the infrared light is from the host
galaxies at redshifts $z \sim 0.7$ \citep{Assefetal2010}. More
significantly, for heavily extincted AGN with E(B$-$V) $= 10$, even modest
amounts of host galaxy contamination ($>10$ per cent) cause the AGN to fall
below the colour cut at $z<1.5$ \citep{Assefetal2010}. Similarly, \cite{Mateosetal2012} found that their selection method completeness is a strong function of AGN luminosity. For AGN with low X-ray luminosity ($L_{2-10 \ keV} < 10^{44}$), the completeness drops from 76.5 per cent to 39.1 per cent for Type 2 AGN. Because of this inability to cleanly identify AGN in the presence of host galaxy contamination using infrared wavelengths alone, other wavelength regimes need to be used in conjunction with WISE to identify AGN.

To this end, we have created a new selection technique that uses the radio continuum emission of AGN to identify the
dustiest AGN with high levels of host galaxy contamination that are not detected by existing WISE colour cuts. The resulting Invisible AGN Catalogue contains galaxies that are ``invisible'' in the Sloan Digital Sky
Survey (SDSS; 355 -- 893 nm; \citealt{Yorketal2000}) and the Two Micron All Sky Survey (2MASS; 1.24 -- 2.16 $\mu$m; \citealt{Skrutskieetal2006}), but are detected in the AllWISE survey (3.4 -- 22
 $\mu$m; \citealt{Cutrietal2014}) and in the Faint Images of the Radio Sky at Twenty-Centimeters
Survey (FIRST; 20 cm / 1.4 GHz; \citealt{Beckeretal1994}; \citealt{Helfandetal2015}). A galaxy's presence in AllWISE but not in
SDSS or 2MASS can indicate that the galaxy is exceptionally dusty, while its
continuum emission at 20 cm can indicate that it contains an AGN. By using a galaxy's radio continuum emission to identify it as an AGN, we have created a way to identify many of the most heavily obscured and host galaxy contaminated AGN that are missed by current WISE colour cuts. 

Unlike many other WISE studies, which require that an object be detected at specific wavelengths, we have kept our selection requirements broad in order to include a wide variety of infrared-bright objects. In addition to finding many of the W1$-$W2 selected AGN discussed above, our catalogue also has the ability to find more extreme objects, such as W1W2-dropouts \citep{Eisenhardtetal2012}, Dust Obscured Galaxies (DOGs; \citealt{Deyetal2008}), and BzK galaxies \citep{Daddietal2004}. W1W2-dropouts, galaxies that are faint or undetected at 3.4 and 4.6 $\mu$m but bright at 12 and 22 $\mu$m, are an extreme class of infrared-bright galaxies whose spectral shapes bear resemblance to high redshift ($z \ge 2$; \citealt{Sternetal2014}) ultra-luminous infrared galaxies (ULIRGs; \citealt{Eisenhardtetal2012}). Approximately half of all currently studied W1W2-dropouts show clear signatures of Type 2 AGN \citep{Bridgeetal2013}, while the lack of a far-infrared dust peak in others suggests dust heating by a heavily embedded AGN, rather than by star formation \citep[e.g.][]{Eisenhardtetal2012, Wuetal2012, Tsaietal2014}.  DOGs, selected based on their low optical to mid-infrared flux ratios, appear to be an optically-faint subset of ULIRGs at $z \sim 2$ \citep[e.g.][]{Houcketal2005, Yanetal2007, Deyetal2008, Lonsdaleetal2009, Donleyetal2010}. Their mid-IR spectral energy distributions
	(SEDs) show that they have similar star formation rates and
	infrared luminosities to sub-millimeter galaxies (SMGs; e.g. \citealt{Bussmannetal2009};
	\citealt{Tyleretal2009}; \citealt{Melbourneetal2012}). 
	Since the launch of WISE, a significantly more luminous type of DOG
	has been discovered, dubbed ``Hot DOGs'' \citep{Wuetal2012}. These
	are W1W2-dropout galaxies that also satisfy the colour cut used to
	select DOGs, but are orders of magnitude more luminous in the
	infrared than the
	typical DOG. Their high dust temperatures indicate heating by
	accretion onto a SMBH \citep{Eisenhardtetal2012, Bridgeetal2013,
	Jonesetal2014, Tsaietal2014, Fanetal2016}. Similar to DOGs and W1W2-dropouts, BzK galaxies, which are selected based on their red optical / near-infrared colours, are active star-forming galaxies at $z > 1.4$ with
	star formation rates $\sim 200$ M$_\odot$ yr$^{-1}$
	\citep{Daddietal2004}.  Although there are definite AGN detections in all of these extreme galaxy classes, the exact fraction that contain AGN remains unclear. By relying on FIRST detections to identify AGN, our selection method can provide an easy way to separate the star-formation dominated and AGN-dominated portions of these classes.

In this paper, we present a new selection method for identifying a large population of infrared-bright, ``invisible'' AGN missed by previous selection methods. This open-ended selection method selects a wide variety of objects, including many that are missed by WISE colour-cuts because of their high levels of host galaxy contamination. Section \ref{method} summarizes the selection method, Section \ref{cataloguestats} presents the Invisible AGN Catalogue and analyzes the effects of the selection method on the parent data, Section \ref{FIRST} explains the identification of AGN using FIRST, Section \ref{discussion} compares the catalogue to other selection methods and discusses the AGN detection rate, and Section \ref{conclusions} summarizes the catalogue's implications. Unless otherwise noted, we use Vega magnitudes and assume $\Omega_M = 0.3$, $\Omega_\Lambda = 0.7$, and $H_0 = 70$ km s$^{-1}$ Mpc$^{-1}$. 

\section{Obscured AGN Selection Method}\label{method}
To create a catalogue of ``invisible'', infrared-bright AGN, we selected all
sources that are detected in the AllWISE and FIRST catalogues, but are not
detected in 2MASS or SDSS Data Releases 7 (DR7; \citealp{Abazajianetal2009}) or 9 (DR9; \citealp{Ahnetal2012}). 
The AllWISE source catalogue \citep{Cutrietal2014} builds upon the initial all-sky catalogue of WISE and contains flux measurements at 3.4, 4.6, 12, and
22 $\mu$m for the entire sky. The primary
advantage of using the AllWISE catalogue over the original WISE catalogue for this work is that AllWISE has
increased sensitivity at 3.4 and 4.6 $\mu$m and improved astrometry over
its predecessor. These improvements are crucial because the selection method is biased towards objects that are faint at all wavelengths (see Sec. \ref{cataloguestats}) and improved astrometry will reduce
the potential false-detection rate in the final catalogue. 

The FIRST catalogue was created using the NRAO Very Large Array and covers over 10,000 square degrees of the Northern and Southern Galactic Caps \citep{Beckeretal1994}. It has a typical flux density rms noise of 0.15 mJy and a resolution of $5''$. It is a radio continuum survey at 20 cm (1.4 GHz).  

SDSS \citep{Eisensteinetal2011} covers a slightly larger area than FIRST and contains over 900 million objects. It has five filters: {\it u, g, r, i,} and {\it z} centred at 3551 \AA, 4686 \AA, 6165 \AA, 7481 \AA, and 8931 \AA, respectively. We primarily use DR9 because it has increased sky coverage (mostly around the Southern Galactic Cap) and improved photometry and sky subtraction over its predecessors. However, DR9 has trouble identifying objects in high background flux areas (typically near a bright star or galaxy), so we also use DR7, which is less affected by this issue. In the catalogue creation, we first remove objects detected in DR9 and then remove any remaining objects that are detected in DR7 (see below).

2MASS is a survey of the whole sky at 1.25, 1.65, and 2.17 $\mu$m \citep{Skrutskieetal2006}. It is a shallow survey (e.g. $>14.3$ mag at 2.17 $\mu$m) generally only capable of detecting large, nearby galaxies that are also detectable by SDSS. Despite this, 2MASS is more accurate at identifying the centres of nearby, extended galaxies than SDSS, making it an important part of removing optically detected objects.

\begin{figure}
\includegraphics[width=\columnwidth]{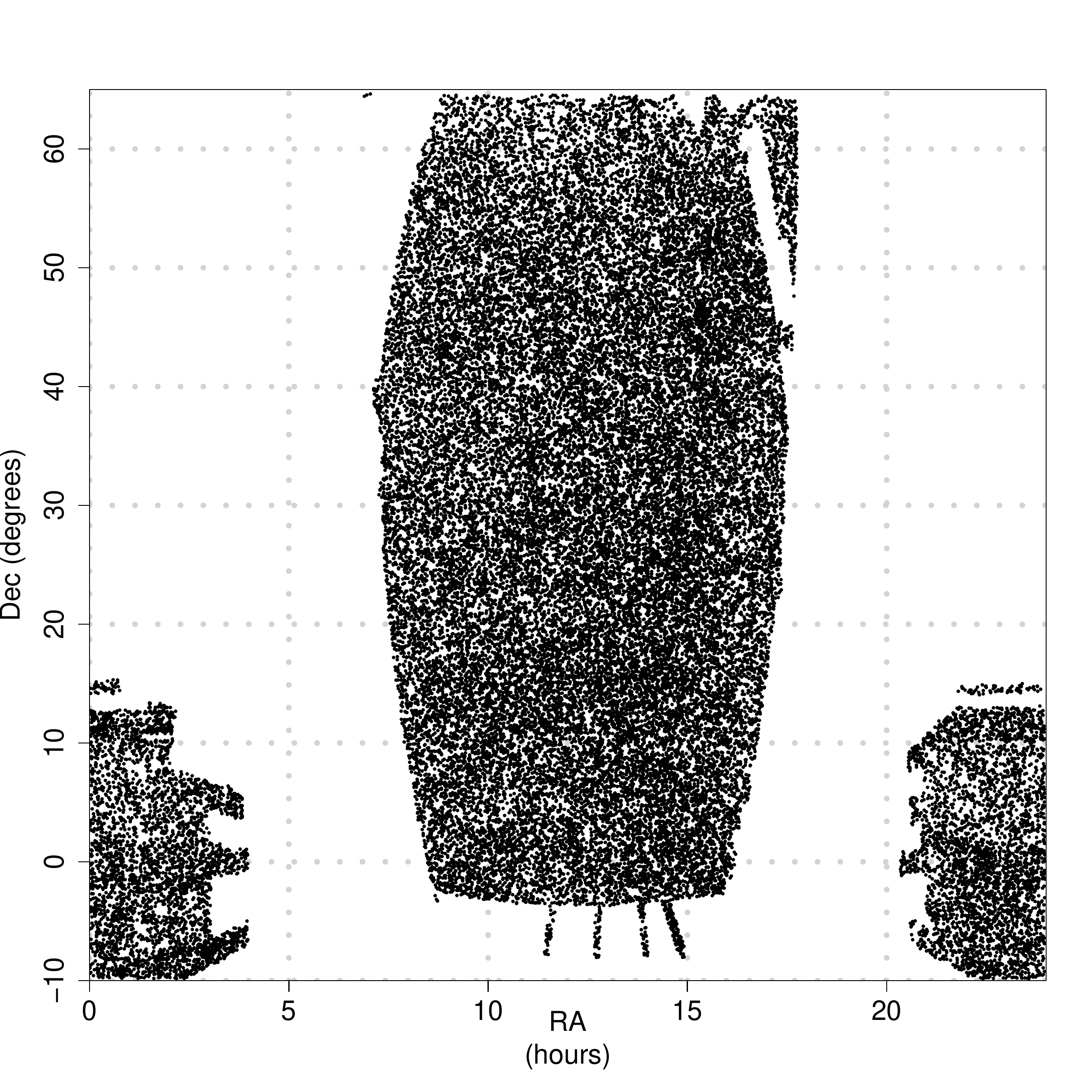}
\caption{Map of all objects included in our final catalogue. The total sky area corresponds to the area covered by FIRST and SDSS DR9.} \label{coverage_map}
\end{figure}

Before creating our catalogue, the FIRST catalogue was pruned to only
include sources that are within the SDSS DR9 survey area. Although FIRST is
designed to cover the same area as SDSS, its footprint does not match
exactly. Because the goal of the catalogue is accuracy rather than
completeness (i.e. it is better to create a smaller catalogue where all
sources are real rather than a catalogue with more sources that are
possible false detections), we adopted a conservative pruning of the FIRST
catalogue. To remove all FIRST points not overlapping with the SDSS DR9
footprint, we created a mesh of points $10'$ apart. All points must have an
SDSS DR9 source within $10'$ of them to be included in this final map. All
edge points were removed to help ensure that catalogue objects are
classified as ``optically-invisible'' because they are not detected by
SDSS, and not because they are outside the SDSS coverage area. Therefore, the final catalogue is likely missing some sources around the edges of the coverage area. Figure \ref{coverage_map} shows a map of the sky area covered by our catalogue. 

We also removed all FIRST objects in areas around bright stars. The presence of a bright star in SDSS increases the average background flux in the surrounding area and often prevents the SDSS catalogue algorithms from effectively identifying nearby sources. To mitigate this effect, we removed a circular region around every star with an {\it r} magnitude $<10$. The radius of the circular region removed scales with the magnitude of the star ($40''$ for $10<r<9$, $60''$ for $9<r<8$, and $120''$ for $r<8$). Finally, we also removed $\sim 50$ individual objects that visual inspection of SDSS images showed to be false detections. All of these individually rejected objects are outliers in flux-flux or color-color space (see Sections 4 and 5).

To construct the catalogue, we first selected all sources within the pruned FIRST catalogue that had an
AllWISE counterpart. In order for a FIRST source to be matched with an AllWISE source, the angular separation between the two sources must be less than the largest 1-sigma position uncertainty among the two catalogues. Typically, FIRST has the largest position uncertainties -- on the order of
$0.9''$.
This technique of tailoring the selection
criterion to each individual source is necessary because the astrometric
precision of AllWISE and FIRST objects vary depending on the magnitude or flux density of the
object. Using this matching criterion, there is only a small probability that an AllWISE source is within $0.935''$ (the average position uncertainty of FIRST) of a FIRST source purely by chance. Based on the source density of AllWISE ($\sim 12400$ -- 14500 sources deg$^{-2}$, depending on galactic latitude) and assuming the source distribution is random, we roughly expect 12400 sources deg$^{-2}$ $\times$ $(0.935'')^2 = 0.0026$ WISE sources within each FIRST source search radius by chance. Thus, we would expect only 0.26 -- 0.31 per cent of FIRST sources to have WISE counterparts purely by chance. This is far less than the actual 5.34 per cent of FIRST sources with WISE counterparts, indicating that the majority of these matches are not random.  

An infrared signal--to--noise ratio (SNR) limit was also placed on the catalogue to exclude possible false detections. For an object to be included in the AllWISE catalogue, it must have a SNR $\ge 5$ in at least one band. However, examination of the AllWISE image database showed that many of these low SNR objects appear to be spurious and should not have been included in AllWISE. Therefore, we adopted a more conservative approach where all WISE sources included in the Invisible AGN Catalogue must have a SNR $> 5$ in at least two bands or a SNR $> 7$ in one band. Ultimately this leads to a catalogue with fidelity and reliability at the cost of completeness. See Section \ref{completeness} for further discussion of catalogue completeness. 

Next, all sources from our catalogue that were detected
in SDSS DR9 or DR7 were removed.
An object has no SDSS counterpart if there are no SDSS
sources within an angular distance equal to three times the object's largest position
uncertainty (either from WISE or FIRST). The position uncertainties of SDSS objects were not considered because these
are much smaller than the uncertainties in WISE and FIRST -- all
SDSS positions are accurate to within $0.1''$
\citep{Pieretal2003}, while FIRST positions are accurate to $1''$ at the survey detection threshold \citep{Whiteetal1997}. Finally, we removed all sources detected in 2MASS in the same way. Once again, 2MASS has much smaller position uncertainties than WISE and FIRST, so they were not considered.

\section{Invisible AGN Catalogue Properties}\label{cataloguestats}
The Invisible AGN Catalogue comprises 46,258 objects. Each catalogue entry contains the object's FIRST and AllWISE fluxes and positions, the
separation between these two positions, and the associated position and separation
uncertainties\footnote{The complete catalogue is available online and at
  http://vizier.u-strasbg.fr/}. Appendix A shows a sample of the first 10 catalogue entries and lists the equations used to calculate source position uncertainty and the angular separation between an object's FIRST and AllWISE positions.
Table \ref{cataloguestats_table} lists several catalogue statistics. There
are 590 objects detected in all four WISE bands, 12,578 detected only in
[3.4], 29,905 detected only in [3.4] and [4.6], 39 detected only in [12]
and [22], and only one detected only in [22]. Additionally,
there are 274 objects with an integrated 1.4 GHz flux density $> 0.1$
Jy and five with a flux density $> 1$ Jy.  The majority of catalogue objects (48,984) have flux densities $< 0.1$ Jy. AllWISE magnitudes are Vega magnitudes calculated from point spread function (PSF) profile fits that are made by chi-squared minimization on a ``stack'' of all single-exposure frames covering a deep source detection \citep{Cutrietal2014}. The majority of our catalogue objects are unresolved so this deep detection profile-fit photometry provides the highest SNR flux measurements with low probability of contamination from other nearby sources. However, 116 Invisible AGN Catalogue objects are marked as extended in the AllWISE catalogue. For these objects, the reported magnitudes may under-represent the total source flux density and we recommend individual examination to ensure accurate flux measurements. We have marked these objects in our catalogue and exclude them from all following plots, unless otherwise indicated.

\begin{table}
 \caption{Median Values of the Invisible AGN Catalogue}
 \tabletypesize{\footnotesize}
 \label{cataloguestats_table}
 \begin{tabular}{lcc}
 \hline
  & Median & 68\% Confidence Interval \\
  \hline
 F$_{\mbox{1.4 GHz}}$ (mJy) & 1.91 & 1.01 -- 5.01 \\ \relax
 [3.4]$ \ ^{\mbox{a}}$ & 16.63 & 16.13 -- 17.13 \\ \relax
 [4.6] & 15.97 & 15.37 -- 16.47 \\ \relax
 [12] & 12.41 & 11.91 -- 12.71 \\ \relax
 [22] & 8.86 & 8.46 -- 9.16 \\
 SNR$_{[3.4]}$ & 12.7 & 8.5 -- 19.1 \\
 SNR$_{[4.6]}$ & 6.6 & 3.9 -- 10.8 \\
 SNR$_{[12]}$ & 0.5 & -0.6 -- 2.1 \\
 SNR$_{[22]}$ & 0.3 & -0.7 -- 1.4 \\
 $\theta$ ($''$)$ \ ^{\mbox{b}}$ & 0.4 & 0.2 -- 0.8 \\
 $\sigma_\theta$ ($''$) & 0.17 & 0.07 -- 0.27  \\
 $\sigma_\alpha^{\mbox{WISE}}$ ($''$) & 0.14 & 0.14 -- 0.24 \\
 $\sigma_\delta^{\mbox{WISE}}$ ($''$) & 0.14 & 0.04 -- 0.24 \\
 $\sigma_\alpha^{\mbox{FIRST}}$ ($''$) & 0.67 & 0.37 -- 1.17 \\
 $\sigma_\delta^{\mbox{FIRST}}$ ($''$) & 0.77 & 0.47 -- 1.17 \\
\hline
 \end{tabular}
 \tablenotetext{a}{AllWISE flux densities are in Vega magnitudes. Sources identified by the AllWISE catalogue as extended are excluded from these calculations.}
 \tablenotetext{b}{The angular separation between the object's AllWISE and FIRST positions.}
\end{table}

\begin{figure}
\includegraphics[width=\columnwidth]{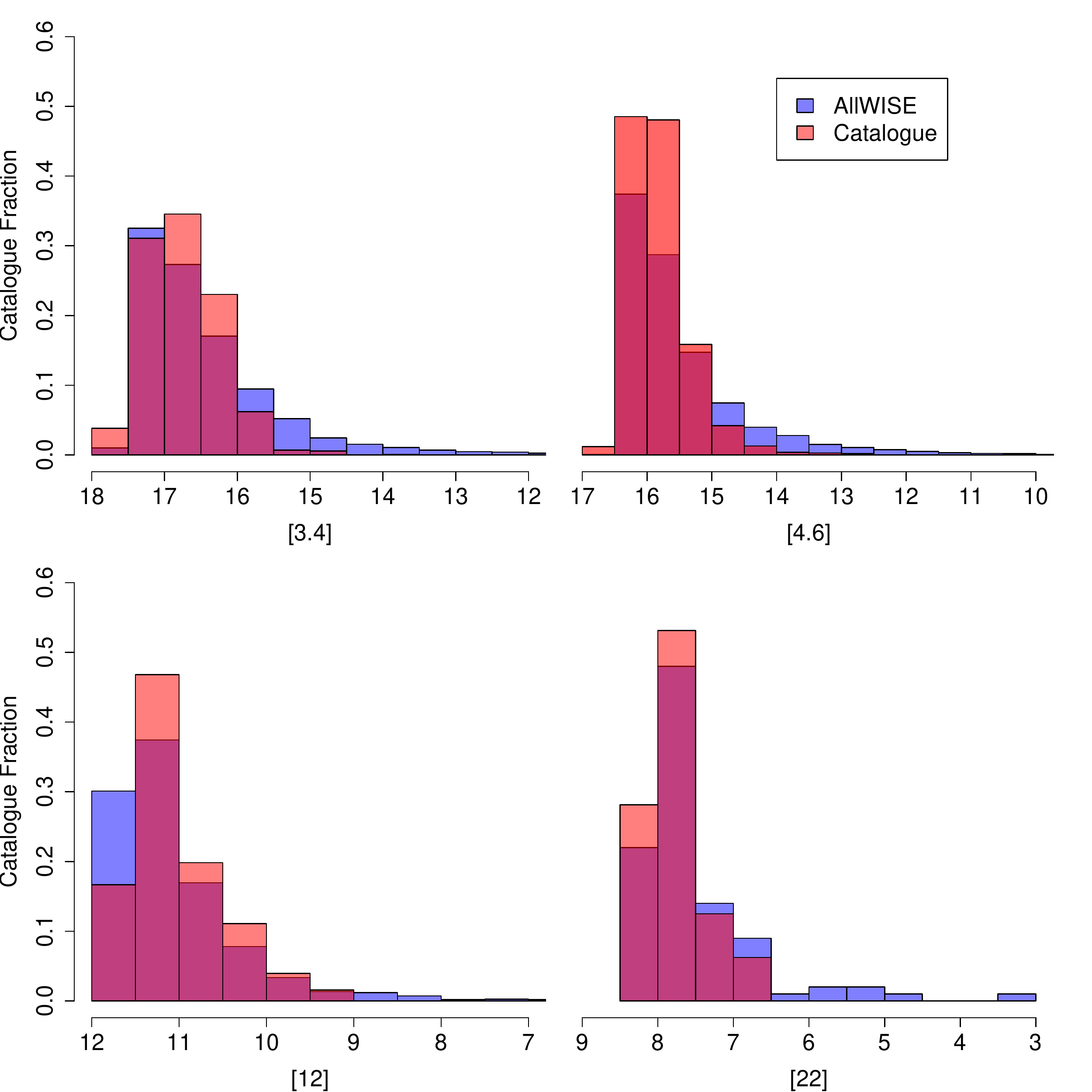}
\caption{Magnitude distributions of the AllWISE bands for our catalogue (salmon) and the full,
unfiltered AllWISE catalogue (blue) both in a $10^\circ$ radius circular region centred at ($12^h, 40^\circ$). The overlap between the catalogues is shown in magenta. Sources identified by AllWISE as extended are excluded from these histograms. \label{WISEdist}} 
\end{figure}

Our selection method biases the resulting catalogue in several
ways. To understand how the WISE sources selected by the Invisible AGN
Catalogue differ from the overall AllWISE catalogue, we compare the
distribution of WISE band magnitudes of our catalogue to the full,
unfiltered AllWISE catalogue in Figure \ref{WISEdist}.  To reduce the
AllWISE catalogue to a computationally manageable size, we restricted our
comparison to a $10^\circ$ radius circular region centered at ($12^h,
40^\circ$). We chose this area because it is centrally located in our
catalogue and contains enough objects on which to perform statistically
robust tests. Two-sample Kolmogorov--Smirnoff (KS) tests for the first three bands show that there is
a $<2$ per cent chance that the AllWISE catalogue and Invisible AGN Catalogue are different in each
band purely by chance. The KS test for [22] showed that the two distributions are statistically the same, however there are few catalogue objects detected in [22] (32 objects) so this KS test is less meaningful. Statistically, the magnitude
distributions in our catalogue are more skewed towards faint, large magnitude
sources than the complete AllWISE catalogue. Individual examination of a
sample of the brightest AllWISE sources ($\sim 100$ objects) in SDSS DR9 shows
that the filtering process removes large, nearby galaxies that are
easily detected at optical wavelengths.

\begin{figure}
\includegraphics[width=\columnwidth]{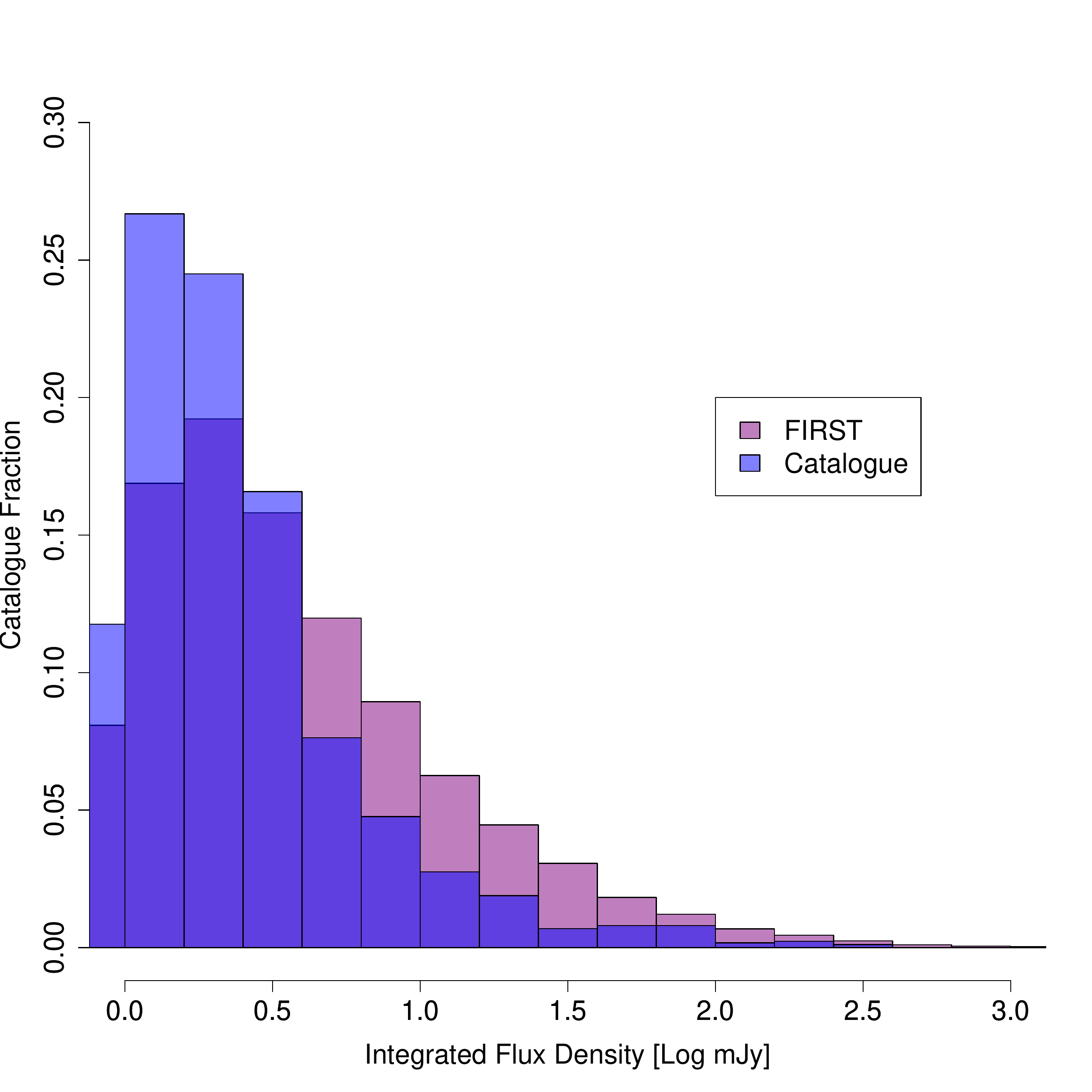}
\caption{FIRST 1.4 GHz integrated flux density distributions for our catalogue (blue) and the full,
unfiltered FIRST catalogue (purple) both in a $10^\circ$ radius circular region centered at ($12^h, 40^\circ$). The overlap is shown in dark violet. \label{Radiodist}} 
\end{figure}

Figure \ref{Radiodist} shows a similar plot to Figure \ref{WISEdist} but
instead compares the distribution of 1.4 GHz FIRST integrated flux densities for our catalogue and the full,
unfiltered FIRST catalogue in a $10^\circ$ radius circular region centered at ($12^h, 40^\circ$) - there are $\sim 29,000$ FIRST sources in this region. A
two-sample KS test shows
that the selection method significantly altered the shape of the radio flux
density distribution ($p < 2.2 \times 10^{-16}$). Once again, our catalogue's distribution skews fainter than the full FIRST catalogue. 

To understand how selecting FIRST objects that are detected in WISE but not
in SDSS alters the distribution of FIRST flux densities in our catalogue,
we created a similar catalogue of objects that are detected in FIRST, but
not in WISE or SDSS. We used all the same processes and selection methods,
but also excluded all FIRST objects that had an AllWISE counterpart. We
found no statistical difference in the distribution of 1.4 GHz flux
densities between the complete FIRST catalogue and this test catalogue of
objects detected in FIRST but not in WISE or SDSS. Thus, we can conclude
that the difference in 1.4 GHz flux density distribution between the Invisible AGN Catalogue and the complete FIRST catalogue is caused by the exclusion of SDSS sources. Visual inspection of a sample of bright 1.4 GHz flux density objects not selected for our catalogue shows that the skewedness shown in Figure \ref{Radiodist} is because many of the brightest FIRST sources originate in dust-poor, optically-bright galaxies that are easily detected by SDSS. Therefore, the skewedness in our catalogue's distribution of FIRST flux densities helps illustrate the success of our selection method; we have biased our catalogue towards dusty, obscured AGN and excluded dust-poor galaxies with unobscured AGN.   

\begin{figure}
\includegraphics[width=\columnwidth]{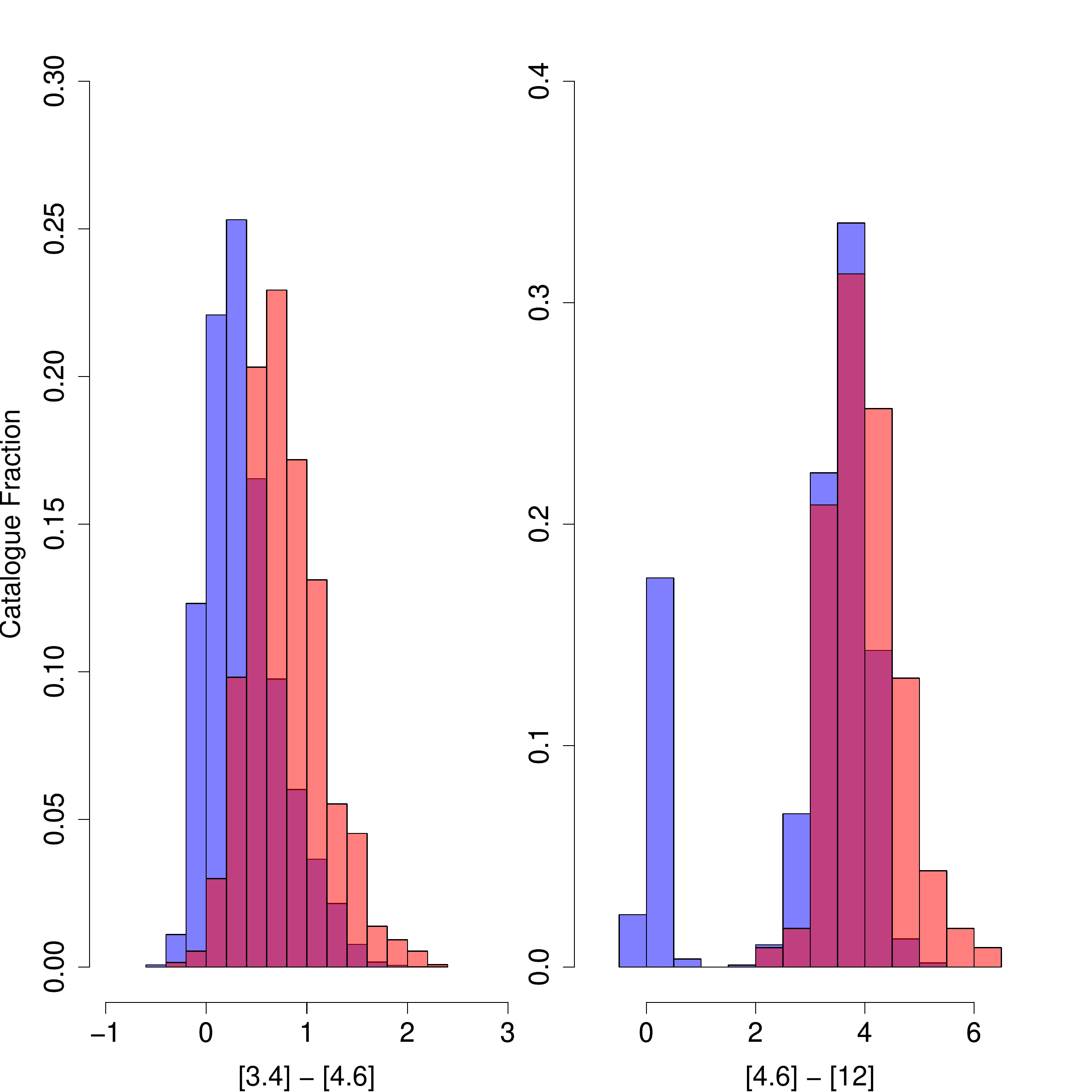}
\caption{AllWISE colour distributions for our catalogue (salmon) and the full,
unfiltered AllWISE catalogue (blue) both in a $10^\circ$ radius circular region centred at ($12^h, 40^\circ$). The overlap is shown in magenta. Sources identified by AllWISE as extended are excluded from these histograms.\label{colourdist}}
\end{figure}

The infrared colours of our catalogue are also altered from the
complete AllWISE catalogue by the selection process (Figure \ref{colourdist}). The KS tests find $p<< 1$ per cent for $[3.4] - [4.6]$ and $[4.6] - [12]$. Most obviously, the selection method removed all
of the $[4.6] - [12] = 0$ sources and most of the $[3.4] - [4.6] = 0$
sources from the filtered catalogue, as well as biasing our catalogue towards redder colours in both cases. Zero-colour sources contain no dust and
are usually stars within our own galaxy or elliptical galaxies (e.g.
\citealp{Wrightetal2010}).  Similarly, red mid-infrared colours typically indicate dust emission. Since the goal of our selection method is to find obscured AGN, this significant reduction in the number of zero-colour sources and the overall shift towards redder colours are excellent indications of the success of the selection method.

\subsection{Completeness}\label{completeness}
The largest constraining factor in our catalogue is the FIRST flux limit. Only $\sim 10 - 20$ per cent of AGN are radio-loud (e.g. \citealt{Kellermannetal1989}; \citealt{Urry&Padovani1995}; \citealt{Sternetal2000}; \citealt{Ivezicetal2002}) and so, by requiring that all catalogue objects have a radio flux $> 1$ mJy, we miss many radio-quiet AGN.
If the average radio-quiet AGN has a 1.4 GHz luminosity $< 10^{25}$ W
Hz$^{-1}$ \citep{Yunetal2001}, FIRST can detect radio-quiet AGN out to $z =
1.3$ (luminosity distance $\sim 9140$ Mpc). All radio-quiet AGN beyond this redshift are likely missed by our selection method.  
However, despite the fact that many radio-quiet AGN are excluded from our catalogue, we are not biasing our catalogue towards the classic radio-loud AGN, which typically resides in a dust-poor elliptical galaxy. In fact, Figures \ref{Radiodist} and \ref{colourdist} show that our catalogue is still biased towards dusty, red galaxies with relatively low radio flux densities. Despite the FIRST flux limit, our other selection criteria (namely the exclusion of galaxies detected in SDSS and 2MASS) have helped to exclude the canonical optically-visible, radio-loud AGN.

Many radio-loud AGN are also omitted by our catalogue because of an offset between the IR and radio emission in many galaxies. The brightest radio sources are typically large jets or lobes of
plasma ejected from a galaxy by the active central black hole \citep{BlandfordRees1974}. In many of these galaxies, the lobe has by far the brightest radio flux, while the
radio core containing the AGN is quite dim in comparison. Because FIRST identifies objects using an automated algorithm that searches for peaks in flux density, it frequently identifies the radio lobe as the object rather than the radio core. After identifying an object, FIRST fits a two dimensional Gaussian to the source in order to estimate its size. However, many of these radio lobes and jets are far from Gaussian shaped and so the size of the object is severely underestimated, making it difficult to automatically identify these lobed objects. In other words, the majority of radio-loud FIRST objects may have WISE counterparts that are missed due to FIRST's detection of radio lobes and jets rather than of the radio core. 
To understand the number of objects potentially missed by this offset, we individually examined a sample of radio-loud FIRST objects. We found that $\sim 30$ per cent of FIRST objects with flux densities $> 300$ mJy do indeed have mid-IR counterparts with no SDSS detection but are not included in our catalogue because their radio flux originates from a jet that is offset from the radio core. This amounts to $\sim 840$ objects, $<< 1$ per cent of the complete FIRST catalogue. 

Our catalogue also includes a number of false detections. To assess the number of objects that should not have been included in our catalogue because they are optically bright enough to be detected by SDSS, we examined the SDSS images of a random sample of 500 catalogue objects. We examined each image by eye and noted any objects that visually appeared to be false detections. Based on this test, we estimate that 3.6 per cent of our objects may be detected in SDSS and thus should not have been included in our catalogue. The most common reasons for a galaxy to be detected in SDSS and erroneously included in our catalogue include poor object centroiding by the SDSS catalogue creation tools, poor object detection because of a nearby bright star that was not removed by our procedures, and small errors in our SDSS coverage map that led to catalogue objects existing outside the footprint of SDSS (we cannot precisely model the sky coverage of SDSS because its edges can vary rapidly and non-linearly as a function of sky position). We have removed objects we have found that lie outside the footprint of SDSS, but more are certainly still present in the final catalogue.

We also performed the same test with a random sample of 500 WISE images of catalogue objects to assess the false detection rate due to objects that are misidentified in WISE. Only seven of the 500 sample objects may be false detections in WISE. Two were on the edge of saturated sources where no second object was visually apparent, while the other five were part of non-Gaussian extended structures where the centre was unclear. With this sample, we can conclude that $\sim 1.4$ percent of the Invisible AGN Catalogue are false-detections due to WISE. 

The AllWISE catalogue includes a contamination / confusion flag to indicate objects that may be false detections due to proximity to bright extended objects, diffraction spikes or halos. In total, 4.6 per cent of the Invisible AGN Catalogue is flagged. However, by examining a random sample of 500 flagged WISE images from our catalogue, we found that the flagging is highly conservative. Only 1.6 per cent of flagged images visually contain false detections -- 0.07 per cent of the AllWISE catalogue. Therefore, we elected to include objects with contamination flags in our final catalogue because exclusion of flagged objects would remove many more real detections than false detections.

Overall, we estimate that $\sim 5$ per cent of our catalogue are false detections. The largest contributor to this false detection rate is inclusion of objects that should be detected in SDSS but do not appear in the SDSS catalogue. Misidentification of objects in WISE also contributes to this rate, but the effect is much less significant.  

\section{Identifying AGN with FIRST}\label{FIRST}
Despite the radio flux limit imposed by the FIRST survey, detection by this
survey is crucial for identifying our objects as AGN. Using the
far-infrared [FIR]--radio relation \citep{PriceDuric1992}, we can separate
star-forming galaxies from AGN-dominated galaxies based on their relative
infrared and 1.4 GHz radio continuum flux densities.  Because FIRST has a
detection limit of 1 mJy, we can use the FIR--radio correlation to place a
clear limit on the distance out to which a star-forming galaxy can be
detected by FIRST (and our catalogue). Based on the FIR--radio relation observed by \cite{Yunetal2001}

\begin{equation}
\log (L_{\mbox{1.4 GHz}}) = \log (L_{60 \ \mu m}/L_\odot) + 12.07
\end{equation}
where $L_{\mbox{1.4 GHz}}$ is in units of W Hz$^{-1}$, and assuming a correlation between the mid-IR flux of a star-forming galaxy and its star formation rate of
\begin{equation}
\mbox{SFR} (M_\odot \mbox{yr}^{-1}) = 1.7 \times 10^{-10} \ L_{8-1000 \ \mu m} (L_\odot)
\end{equation}
\citep{Kennicutt1998}, where the ratio of $L_{8-1000 \ \mu m}$ to $L_{60 \ \mu m}$ is observed to be 2.663 for a normal star-forming galaxy \citep{Yunetal2001}, we derive a relationship between the observed radio flux density of a star-forming galaxy, its star formation rate, and its luminosity distance:
\begin{equation}
S_{\mbox{1.4 GHz}} = 21.59 \ \frac{\mbox{SFR}}{D_L^2}
\end{equation}
where $S_{\mbox{1.4 GHz}}$ is in Jy, SFR is in M$_\odot$ yr$^{-1}$ and $D_L$ is
in Mpc. A galaxy with a modest SFR of 1 M$_\odot$ yr$^{-1}$ can only be
detected out to $\sim 146$ Mpc ($z \sim 0.033$) at the FIRST detection
limit (1 mJy). A more extreme object with a high star formation rate of 100
M$_\odot$ yr$^{-1}$ can be detected out to a distance of $\sim 1470$ Mpc,
equivalent to a redshift of $\sim 0.29$. Therefore, FIRST is able to detect
average star-forming galaxies at $z < 0.033$ and most other galaxies with
high star-formation rates at $z < 0.29$. In other words, the majority of
the Invisible AGN Catalogue objects with $z > 0.29$ must contain AGN in
order to be detected by FIRST and thus be included in the catalogue. For
comparison, the
FIRST Bright Quasar Survey \citep{Whiteetal2000} of optically bright
quasars detected in FIRST has a mean redshift of $\sim 1.05$. If optically-faint FIRST quasars have a similar redshift distribution, then the majority of our catalogue should lie above $z = 0.29$ and can be classified as AGN rather than star-forming galaxies.

\begin{figure}
\includegraphics[width=\columnwidth]{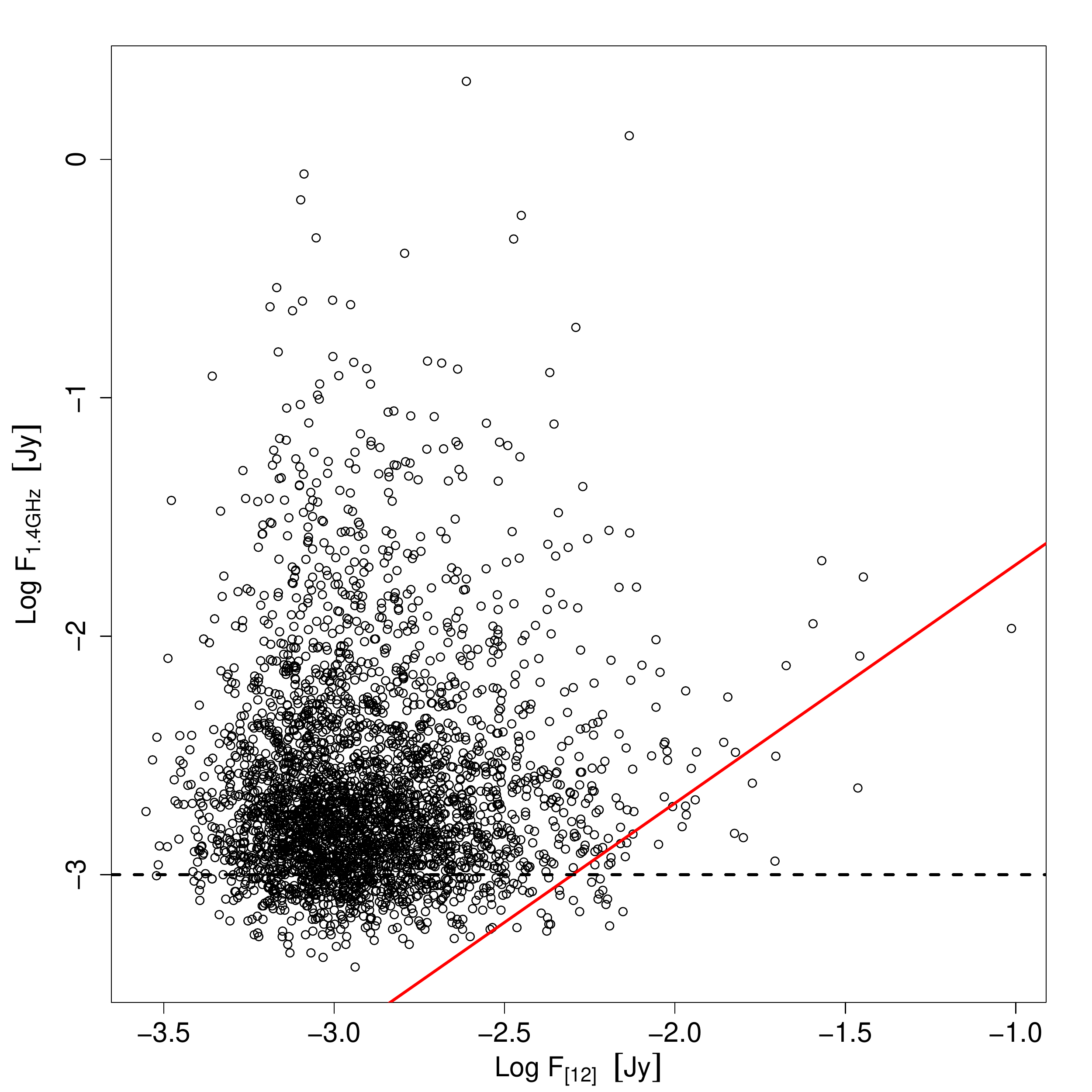}
\caption{WISE 12 micron vs 1.4 GHz integrated flux densities for all objects detected at 12 microns. The red line represents the linear relationship between FIR and radio fluxes for galaxies whose radio continuum is dominated by free-free emission from star-formation. The dashed line is the 1 mJy sensitivity limit of the FIRST survey. Some points lie below this limit because FIRST assumes all sources are Gaussian when calculating the integrated flux. This can sometimes lead to an underestimation of the integrated flux density. \label{FIRRadio}} 
\end{figure}

The FIR--radio relation can be seen in our catalogue if we do not exclude
objects detected in 2MASS (Figure \ref{FIRRadio}). As shown in
\cite{Yunetal2001}, the star-forming galaxies have high 12 micron flux
densities and follow the expected linear relationship, while the AGN have
lower 12 micron flux densities and show no significant correlation. To
guide the eye we included a line with $q=1$, the expected FIR--radio ratio
for star-forming galaxies. As an aside, this plot demonstrates the importance of excluding 2MASS detected objects from our catalogue; without this exclusion our catalogue would still contain many star-forming galaxies without AGN.

\begin{figure}
\includegraphics[width=\columnwidth]{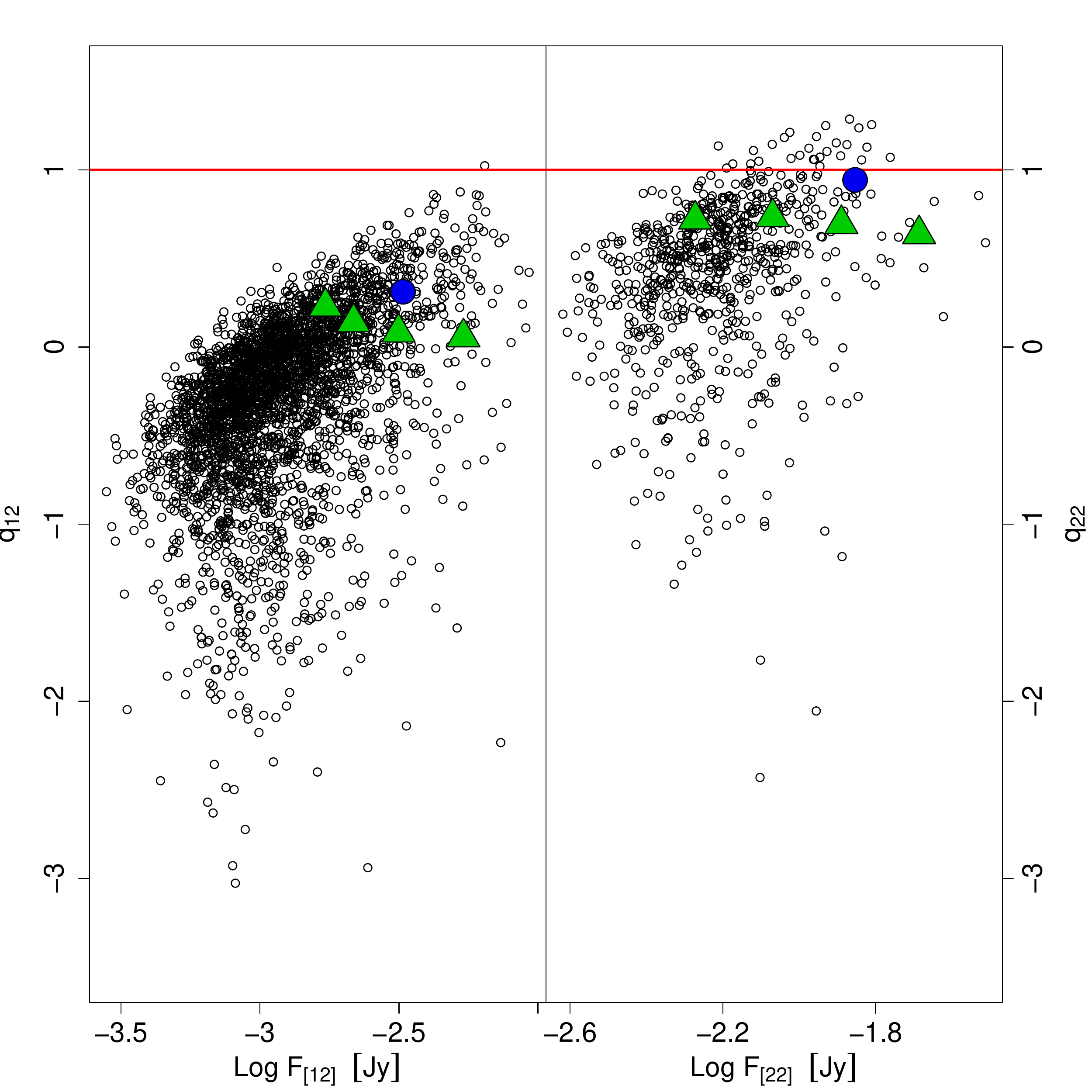}
\caption{IR--radio ratio for all points detected at 12 and 22 microns, respectively, plotted against their 12 or 22 micron flux densities. The red line represents the $q$ value expected for galaxies without AGN whose radio emission is powered only by star formation. The $q$ values of Arp 220 at $z=0.2$ (blue circle) and an average of 17 SMGs at $z=0.4 - 0.7$ (green triangles, redshift increases by $\Delta z = 0.1$ to the right) are plotted to show the lower extreme $q$ values possible for star-formation dominated galaxies. The sharp diagonal edge of the data is caused by the 1 mJy detection limit of FIRST. \label{q}} 
\end{figure}

Figure \ref{q} shows the FIR--radio ratio for all Invisible AGN Catalogue sources detected at either 12 or 22 microns, where $q_{12} = \log_{10} \left (F_{[12]} \ / \ F_{1.4 \ GHz} \right)$ and $q_{22} = \log_{10} \left (F_{[22]} \ / \ F_{1.4 \ GHz} \right)$. 
Most objects with a $q<1$ at either wavelength have more radio emission than can reasonably be produced by star formation alone -- they most likely contain an AGN. However, scatter around the FIR--radio relation (seen in Fig. \ref{FIRRadio}) means that some extreme star-forming galaxies will have $q<1$ even though the majority of their radio flux is created by star formation. As examples of this, Figure \ref{q} also includes the $q$ values of Arp 220 (a canonical ULIRG with a heavily obscured AGN) at $z=0.2$ and an average SMG at $z=0.4, \ 0.5, \ 0.6,$ and $0.7$. These values were computed from theoretical SEDs obtained from \cite{CharyElbaz2001} and \cite{Popeetal2008}, respectively. Arp 220 and the SMG may contain heavily obscured AGN, but the majority of their luminosity is most likely powered by star formation \citep[e.g.][]{Spoonetal2004, Popeetal2008, Smolvicetal2015}. These are extreme cases of star-forming galaxies and thus their $q$ values demonstrate the lowest possibles $q$ values that can be expected from galaxies whose radio emission is dominated by star formation. Additionally, SMGs and ULIRGs are relatively rare, and so, even though many of our objects have $q$ values above those for Arp 220 and the average SMG, we still believe that most are AGN-dominated. Overall, Figure \ref{q} shows that most of the Invisible AGN Catalogue objects with 12 or 22 micron detections contain AGN, especially those with $q$ values below Arp 220 and the average SMG.

\section{Discussion}\label{discussion}
\begin{figure}
\includegraphics[width=\columnwidth]{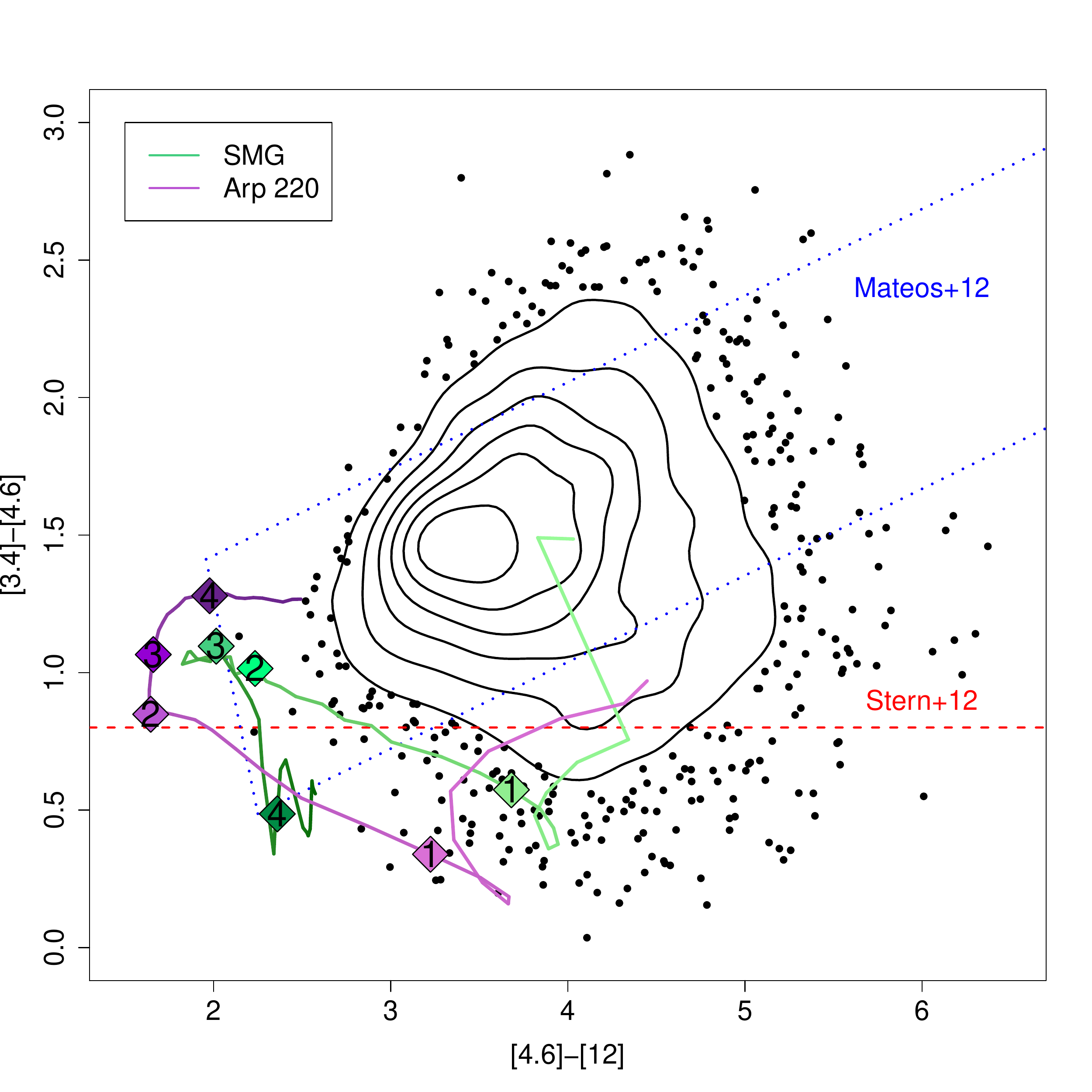}
\caption{WISE colour -- colour plot for all catalogue objects
 detected in the first three WISE bands. Zero on the axes corresponds to
 a stellar mid-IR colour, while higher numbers tend to represent redder,
 dustier objects \citep[e.g.][]{Sternetal2005}. The contours show the probability density of our objects, starting at $p=0.6$ and decreasing by $\Delta p=0.1$ as one moves outwards. The points represent the $p<0.1$ outliers. The red dashed line
 represents the AGN colour selection method used by \citet{Sternetal2012} and the dashed blue lines show the AGN selection method defined by \citet{Mateosetal2012}. All points above the red line and/or within the blue shape are identified as AGN by these methods. Also plotted are the colours of Arp 220 and an average of 17 SMGs between $0<z<5$ (the line colours darken with increasing redshift and the diamonds indicate integer redshift values).} \label{colourcolour} 
\end{figure}

\begin{figure}
\includegraphics[width=\columnwidth]{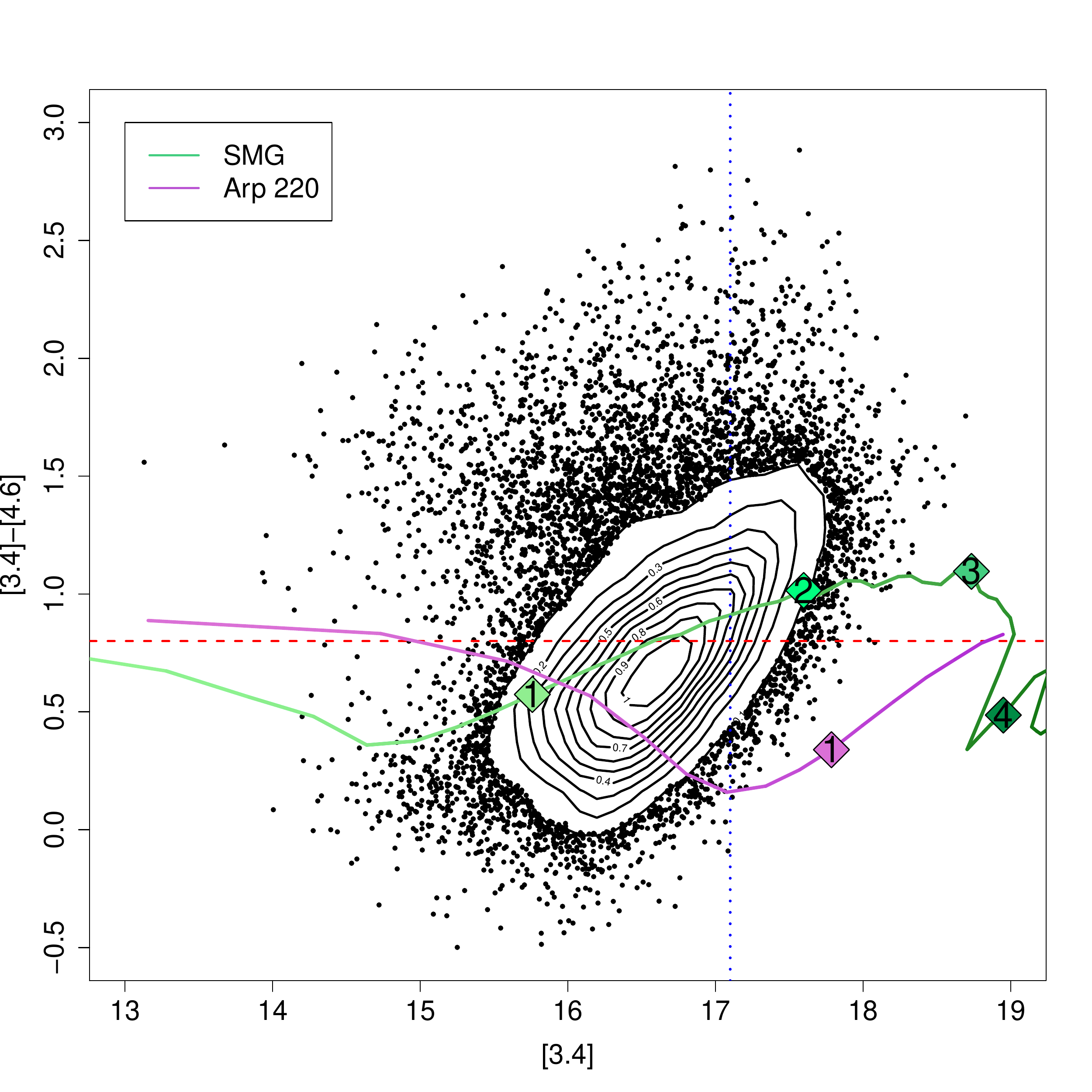}
\caption{WISE [3.4] $-$ [4.6] colour vs [3.4] magnitude for all catalogue objects
 detected in the first two WISE bands (SNR $>5$ in both bands). The dashed horizontal line shows the \citet{Sternetal2012} colour cut. The contours show the probability density of
  our objects, starting at $p=1$ and decreasing by $\Delta p=0.1$ as one moves outwards. The points represent the $p<0.1$ outliers. The colour vs. [3.4] of Arp 220 and the average SMG are plotted between $0.1<z<1.6$ and $0.4<z<4$, respectively. The diamonds indicate integer redshift values. The vertical line shows the 95 per cent complete magnitude limit for [3.4]. The WISE magnitude limits vary as a function of sky position due to variations in survey coverage, background flux, and source density, thereby enabling many objects to be detected below this limit.} \label{w1w2} 
\end{figure}

As discussed in Section \ref{intro}, several IR colour cuts have been used
 in previous studies to select AGN from the WISE survey. Figure
 \ref{colourcolour} shows a colour -- colour plot of all catalogue objects
 detected in the first three WISE bands, along with the W1 $-$ W2 $\geq
 0.8$ AGN selection criterion developed by \cite{Sternetal2012} and the
 colour selection wedge created by \cite{Mateosetal2012}. To be identified
 as an AGN by these methods, an object must fall either above the W1 $-$ W2
 line or within the wedge, respectively. As another way of viewing our
 sample, Figure \ref{w1w2} shows a colour -- magnitude plot of all
 catalogue objects detected in the first two WISE bands, along with the
 same \cite{Sternetal2012} colour cut. 57.4 per cent of our catalogue
 objects that are eligible for the W1 $-$ W2 colour cut (i.e. are detected
 with SNR $>5$ at 3.4 and 4.6 microns) would not be identified as AGN by
 \cite{Sternetal2012} and 25.5 per cent of eligible objects would not be
 selected as AGN by \cite{Mateosetal2012}. This is a good
   indication that our selection method has identified the more obscured AGN
   with high levels of host galaxy contamination that are missed by these
   selection methods. Additionally, both of these selection methods are luminosity dependent; at low AGN luminosities, the colour cuts suffer from contamination by star-forming galaxies. The \cite{Mateosetal2012} wedge is 76.5 per cent complete for Type 2 AGN with $L_{2-10 \ keV} > 10^{44}$ erg s$^{-1}$, but only 39.1 per cent complete for Type 2 AGN with $L_{2-10 \ keV} < 10^{44}$ erg s$^{-1}$. The \cite{Sternetal2012} colour cut is 78 per cent complete for all AGN, but also suffers increased contamination at low AGN luminosities. By requiring a detection by FIRST, our selection method provides a means to separate low-luminosity AGN from star-forming galaxies.  
 
 Also plotted in Figures \ref{colourcolour} and \ref{w1w2} are the colours of Arp 220 and the same average SMG model used in Figure \ref{q} at a range of redshifts between $0<z<5$ -- note that Arp 220 and the average SMG are not detected by FIRST out to $z\sim 5$, but we include their $z< 5$ redshift evolution for illustrative purposes. The majority of the light in the Arp 220 and average SMG templates is created by star formation in the host galaxies. Arp 220 is known to contain a heavily obscured AGN, while it is uncertain whether the SMGs used to create the average model contain AGN. The \cite{Sternetal2012} colour cut selects the SMG template at $z<0.1$ and $1.4 < z < 3.6$, and the Arp 220 template at $z \sim 0.2$ and $1.6<z<5$. The \cite{Mateosetal2012} wedge selects the SMG template at $z<0.1 $, $1.3<z<2.2$, $3.1<z<3.7$, $4.1<z<4.2$, and $4.7 < z <4.8$, and the Arp 220 template at $z \sim 1.3$ and $z > 4.1$. Thus, there are many redshifts at which the two templates are selected by neither the W1 -- W2 colour cut nor the wedge, demonstrating how high levels of host galaxy light can confuse these selection methods -- ideally, the galaxies would stay on one side of the selection criterion, regardless of redshift.  Without the addition of a FIRST selection criterion, WISE selection methods are unable to cleanly separate star-forming galaxies from low-luminosity AGN.

Using a comparison of WISE, FIRST, and X-ray to identify AGN and differentiate them from star-forming galaxies, \cite{Mingoetal2016} also concluded that the \cite{Sternetal2012} and \cite{Mateosetal2012} colour cuts are incomplete.  They found that they are biased against low-luminosity AGN both because the colour cuts are necessarily conservative in order to create clean rather than complete samples of AGN and because W3 is insensitive and unable to detect faint AGN at redshifts beyond $\sim 0.1$. We reach a similar conclusion and present our selection method as a way to define a cleaner sample of low-luminosity AGN. \cite{Mateosetal2012} find that purely star-forming galaxies contaminate their selection wedge at $z>1.3$. In our catalogue, all objects with $z>0.29$ must be AGN in order to be detected by FIRST. Thus, by using the FIR--radio relation to identify AGN, we have created a selection method that can cleanly detect AGN with high levels of dust extinction and host galaxy contamination, even at high redshifts. 

There are several reasons a galaxy with an AGN might not be selected by the W1 $-$ W2 colour cut. For a source with pure, unobscured AGN emission and no galaxy contribution, the source will fall below the colour cut at $z \geq 3.4$ due to the decrease in AGN-heated dust emission at blue wavelengths and to H$\alpha$ emission shifting into the 3.4 micron band \citep{Assefetal2010}. However, even modest amounts of dust extinction (E(B$-$V)$ > 0.1$) significantly redden AGN at all redshifts, bringing them back above the colour cut. Since it is unlikely to find an AGN with no dust extinction or host galaxy contamination, we do not believe high redshift is the sole reason many of our sources are not selected by the \cite{Sternetal2012} colour cut.

A more realistic reason many of our AGN are below the colour cut is because of a combination of host galaxy contamination and dust extinction. Even with modest extinction of E(B$-$V) $ = 1$, the W1 $-$ W2 colour cut is only able to select galaxies where the AGN contributes $\geq 80$ per cent of the mid-infrared emission (see Fig. 2 in \citealp{Sternetal2012}). Assuming $R_V \sim 3.1$ \citep[e.g.][]{Cardellietal1989, GordonClayton1998, Yorketal2006} and a SMC-like gas-to-dust ratio ($N_H / A(V) \sim 2 \times 10^{22}$ cm$^{-2}$ mag$^{-1}$; \citealp{Maiolinoetal2001}), E(B$-$V) $ = 1$ corresponds to $N_H \sim 6 \times 10^{22}$ cm$^{-2}$. For heavily extincted AGN with E(B$-$V) $=10$ ($N_H \sim 6 \times 10^{23}$ cm$^{-2}$), essentially all AGN at redshifts greater than a few tenths are missed by the colour cut, regardless of the level of host galaxy contamination. Arp 220, plotted in Figures \ref{colourcolour} and \ref{w1w2}, provides an example of the many redshifts at which a heavily obscured AGN in a bright host galaxy can be missed. Like Arp 220, we believe that the majority of our objects that fall below the W1 $-$ W2 colour cut are AGN with $N_H > 6 \times 10^{22}$ cm$^{-2}$ whose host galaxies account for at least 20 per cent of their overall mid-infrared light.   

At first glance it appears that the \cite{Mateosetal2012} wedge is more efficient at detecting our objects than the \cite{Sternetal2012} W1 $-$ W2 colour cut. However, the major disadvantage of the wedge is that it requires a detection in three WISE bands, especially at [12]. The [12] band is much less sensitive than the 3.4 and 4.6 micron bands, so including it in a colour cut significantly reduces the number of objects that are eligible for selection. In fact, 94 per cent of our catalogue is not eligible because the objects are not significantly detected in all three of the required bands. The W1 $-$ W2 colour cut is more inclusive, but 29 per cent of our catalogue ($\sim 13,000$ objects) is still not eligible for this selection method. 52 of these qualify as W1W2-dropouts, indicating that they are dusty, ULIRG-like objects at $z>1.6$ \citep{Bridgeetal2013}. Additionally, 639 objects qualify as DOGs, a population of dust enshrouded objects at a redshift range of $1.5<z<2.5$ \citep{Daddietal2004}. 

By not imposing any colour cuts on our sample, we are able to easily identify a large number of these extreme, infrared-bright galaxies. Given the redshift ranges of these object classes ($z>1.5$), we can also conclude that these sources contain AGN; FIRST can only detect galaxies at $z>0.29$ if they contain AGN (Sec. \ref{FIRRadio}). However, by requiring a FIRST detection, we have also limited ourselves to only the radio-loud portions of these populations -- at $z>1.3$ FIRST only detects radio-loud AGN (Sec. \ref{FIRRadio}). In previous studies, it has been unclear what fraction of W1W2-dropouts and DOGs contain AGN. \cite{Deyetal2008} found spectral energy distributions from DOGs indicative of obscured AGN, but no conclusive way to separate the AGN-dominated galaxies from the star-formation dominated galaxies. By combining our FIRST detection requirement with these dusty, infrared-bright galaxy selection techniques, we have found a more efficient way to identify the radio-loud, AGN-dominated portions of these extreme galaxy classes.

\section{Conclusions}\label{conclusions}
We have created a new method for selecting obscured AGN that is unbiased towards heavily obscured AGN and AGN with high levels of host galaxy contamination. We selected objects that are detected in AllWISE and FIRST, but are not detected in SDSS or 2MASS. By excluding all optically detected sources and requiring a detection in AllWISE, we have biased our catalogue towards the most heavily obscured AGN, which are also the most frequently missed by other selection methods. By including a FIRST requirement, we have introduced an easier, cleaner way to identify AGN that does not rely on WISE colour cuts. FIRST is only able to detect all but the most extreme star-forming galaxies (SFR $> 100$ M$_\odot$yr$^{-1}$) out to a redshift of $z\sim 0.29$. Since the majority of $z<0.29$ galaxies are easily seen in the optical and/or near-IR, selecting objects that are detected in FIRST but not SDSS/2MASS ensures that the majority of detected galaxies contain AGN.

Overall, our method selected 46,258 obscured AGN candidates, many of which
are also selected as AGN by other methods. 30.0 per cent of our objects are
AGN using the \cite{Sternetal2012} W1 $-$ W2 colour cut and 4.4 per cent are AGN using the W1 $-$ W2 $-$ W3 \cite{Mateosetal2012} wedge. 6.7 per cent of our objects are AGN based on their mid-IR to
radio flux ratios.  These fractions are low because all of these selection techniques require detection in multiple WISE bands. The \cite{Mateosetal2012} wedge and FIR--radio relation in particular require detection in the less sensitive W3 and/or W4.  Accounting for overlap between these three selection methods, $\sim $ 30 per cent of our catalogue are identified as AGN by these techniques. The remaining 70 per cent represent a potential
previously-unidentified population of candidate AGN.

For those AGN that are detected in the first two WISE bands and can therefore be selected based on their W1 $-$ W2 colour, the population that lies below the \cite{Sternetal2012} colour cut may in part represent highly obscured AGN with host galaxy contamination. Previous selection methods have been unable to cleanly select these heavily embedded AGN because they are unable to differentiate between low-luminosity AGN emission and emission from star formation. However, by selecting galaxies detected in FIRST, we are able to disentangle these two emission sources and select a larger population of heavily obscured AGN that are missed by standard WISE colour cuts. 

Without redshifts or any other additional information about our selected population, it is impossible to make a reliable calculation of our selection method's AGN detection rate. Based on previously developed WISE colour cuts and the FIR--radio relation, we estimate that $>$ 30 per cent of our catalogue are AGN. However, it is the remaining 70 per cent that are the real success of this detection method. These are the candidate AGN that are missed in previous selection methods, most likely due to their high levels of dust obscuration, their high redshifts, and/or their high levels of host galaxy contamination. Based on their exclusion from the \cite{Sternetal2012} colour cut, we estimate that many of these AGN have dust extinction of E(B$-$V) $>1$ ($N_H > 6 \times 10^{22}$ cm$^{-2}$) and host galaxies that contribute at least 20 per cent of the total infrared flux. We also find that $\sim 700$ objects qualify as either W1W2-dropouts and / or DOGs. Based on the estimated redshift distribution of these galaxy classes and their detection by FIRST, we conclude that these are another $\sim 700$ radio-loud AGN found by our selection method.  

Obscured AGN are predicted to outnumber unobscured AGN by a factor of three
and are required to explain the observed infrared and X-ray
backgrounds. However, many of these obscured AGN are missed by existing
WISE selection methods because of the difficulty of differentiating infrared
emission from star formation in the host galaxy and emission from dust
heated by the AGN. With the Invisible AGN Catalogue, we demonstrate a new
method of identifying AGN, regardless of their level of dust obscuration or
host galaxy contamination. By selecting galaxies detected by FIRST and
using the FIR--radio relation, we have demonstrated a new AGN selection
method that can differentiate between these two emission mechanisms without
excluding low-luminosity, obscured AGN. With this new selection method, it will be much simpler to identifying the remaining missing predicted population of obscured AGN.  

\section*{Acknowledgments}
The authors thank the anonymous referee for helpful comments, Mark Taylor for answering many questions about STILTS,
Alexandra Pope for providing the SMG SED template, and Julie Comerford,
Nils Halverson, Dave Brain and John Stocke for useful discussions. All
catalogue comparisons were performed using the tabular manipulation tool
STILTS \citep{Taylor2006}. 

Funding for SDSS-III has been provided by the Alfred P. Sloan Foundation,
the Participating Institutions, the National Science Foundation, and the
U.S. Department of Energy Office of Science. The SDSS-III web site is
http://www.sdss3.org/. SDSS-III is managed by the Astrophysical Research
Consortium for the Participating Institutions of the SDSS-III Collaboration
including the University of Arizona, the Brazilian Participation Group,
Brookhaven National Laboratory, Carnegie Mellon University, University of
Florida, the French Participation Group, the German Participation Group,
Harvard University, the Instituto de Astrofisica de Canarias, the Michigan
State/Notre Dame/JINA Participation Group, Johns Hopkins University,
Lawrence Berkeley National Laboratory, Max Planck Institute for
Astrophysics, Max Planck Institute for Extraterrestrial Physics, New Mexico
State University, New York University, Ohio State University, Pennsylvania
State University, University of Portsmouth, Princeton University, the
Spanish Participation Group, University of Tokyo, University of Utah,
Vanderbilt University, University of Virginia, University of Washington,
and Yale University. 

This publication makes use of data products from the Wide-field Infrared
Survey Explorer, which is a joint project of the University of California,
Los Angeles, and the Jet Propulsion Laboratory/California Institute of
Technology, and NEOWISE, which is a project of the Jet Propulsion
Laboratory/California Institute of Technology. WISE and NEOWISE are funded
by the National Aeronautics and Space Administration. 

This publication makes use of data products from the Two Micron All Sky
Survey, which is a joint project of the University of Massachusetts and the
Infrared Processing and Analysis Center/California Institute of Technology,
funded by the National Aeronautics and Space Administration and the
National Science Foundation.

The National Radio Astronomy Observatory is a facility of the National
Science Foundation operated under cooperative agreement by the Associated
Universities, Inc.



\bibliographystyle{mnras}
\bibliography{wisefirst_catalog}


\newpage

\appendix

\section{The Invisible AGN Catalogue}
\begin{landscape}

\begin{deluxetable}{cccccccccccclclclclccc}
\tablecolumns{22}
\tablecaption{The Invisible AGN Catalogue}
\tablewidth{0pc}
\setlength{\tabcolsep}{5pt}
\tabletypesize{\scriptsize}
\tablehead{
\multicolumn{7}{c}{FIRST Catalogue} & \multicolumn{13}{c}{AllWISE Catalogue} & \colhead{} & \colhead{} \\
\colhead{FIRST ID}  & \colhead{RA} & \colhead{$\sigma$} & \colhead{DE} &\colhead{$\sigma$} &  \colhead{F$_{1.4 \ GHz}$} & \colhead{Sky RMS} & \colhead{AllWISE ID} & \colhead{RA} &\colhead{$\sigma$} &\colhead{DE} & \colhead{$\sigma$} & \colhead{[3.4]} & \colhead{SNR} &\colhead{[4.6]} & \colhead{SNR} & \colhead{[12]} & \colhead{SNR} & \colhead{[22]} & \colhead{SNR} & \colhead{$\theta$} & \colhead{$\sigma_\theta$} \\
\colhead{} & \colhead{deg} & \colhead{$''$} & \colhead{deg} & \colhead{$''$} & \colhead{mJy} & \colhead{mJy/beam} & \colhead{} & \colhead{deg} & \colhead{$''$} & \colhead{deg} & \colhead{$''$} & \colhead{} & \colhead{} & \colhead{} & \colhead{} & \colhead{} & \colhead{} & \colhead{} & \colhead{} & \colhead{$''$} & \colhead{$''$}
}
\startdata
J000043.7-085635 & 0.182421  & 0.86   &  -8.943325  & 0.86   &  1.78 &  0.155  & J000043.76-085636.1 &     0.1823589  & 0.1979   &  -8.9433670 &  0.1901&   16.421 (0.075)&   14.4  & 15.089 (0.108) &  10.1   & 11.406 (0.231)  &  4.7    & 8.666       &      0.4  &  0.27 &  0.23 \\
J000044.3+030754    &  0.184742 &  0.29    &  3.131786&   0.35  &  46.61 &  0.134  & J000044.33+030754.1   &   0.1847368  & 0.0962  &    3.1317018 &  0.0929 &  15.389 (0.043) &  25.1 &  14.210 (0.049)&   22.3 &   10.827 (0.106) &  10.2   &  8.441          &   1.3 &   0.30 &  0.09 \\
J000051.9+030058  &    0.216479 &  1.06    &  3.016144 &  0.76   &  1.00  & 0.136 &  J000051.88+030058.2     & 0.2161902 &  0.3610  &    3.0161766 &  0.3624 &  16.909 (0.116)&    9.3  & 16.334   &         1.9   & 11.749   &         1.7  &   8.874     &        0.1  &  1.05&   0.21 \\
J000055.5+092654 &     0.231383 &  0.65   &   9.448397  & 0.49  &   3.47&   0.118 &  J000055.55+092654.1   &   0.2314717 &  0.1965     & 9.4483837  & 0.1886  & 16.188 (0.068)  & 16.0 &  16.058 (0.193) &   5.6   & 12.227     &       0.9 &    9.095     &        0.0   & 0.32  & 0.14 \\
J000103.6+065659     & 0.265117  & 0.40    &  6.949833  & 0.43  &   4.74 &  0.126&   J000103.64+065659.4 &     0.2651743 &  0.1376  &    6.9498453  & 0.1371 &  15.813 (0.051)&   21.4 &  15.184 (0.095) &  11.5 &   12.054         &   1.1  &   8.995    &        -0.3  &  0.21 &  0.12 \\
J000108.7+032724  &    0.286517  & 0.38    &  3.456881  & 0.42  &   7.14&   0.137  & J000108.76+032724.8     & 0.2865319 &  0.1512   &  3.4568928  & 0.1450  & 15.824 (0.053) &  20.3  & 15.578 (0.127)  &  8.5 &   12.549  &         -0.3  &   8.394    &         1.3   & 0.07  & 0.12 \\
J000110.4-035719     & 0.293400  & 1.50    & -3.955400  & 1.44    & 1.60 &  0.156  & J000110.47-035719.3     & 0.2936625 &  0.2531 &    -3.9553667  & 0.2414 &  16.504 (0.085)&   12.8  & 15.939 (0.192)  &  5.7 &   12.215   &         0.4  &   8.349       &      1.5 &   0.95 &  0.37 \\
J000110.4-092234  &    0.293667  & 0.96   &  -9.376136 &  0.94    & 1.13 &  0.145&   J000110.50-092233.7   &   0.2937551 &  0.1922  &   -9.3760519  & 0.1836 &  16.261 (0.072)   &15.0  & 15.238 (0.107) &  10.1&    12.462     &      0.1   &  8.991   &         -0.5   & 0.44  & 0.25 \\
J000112.4+003554  &    0.301992&   0.87    &  0.598556 &  1.26   &  1.11 &  0.100  & J000112.48+003554.2    &  0.3020375  & 0.3268   &   0.5983957 &  0.3099  & 16.854 (0.119) &   9.1 &  15.889 (0.167)  &  6.5&   12.604     &      -0.1 &    8.569      &       0.8  &  0.60 &  0.33 \\
J000114.5+031740  &    0.310429 &  1.07   &   3.294642  & 0.96  &   1.54 &  0.123 &  J000114.48+031740.1    &  0.3103449  & 0.2097 &     3.2944881  & 0.2017 &  16.233 (0.071)&   15.3&   15.960 (0.194) &   5.6   & 11.865     &       1.5   &  8.367   &          1.0   & 0.63&   0.25\\

\enddata
\tablenotetext{a}{The first 10 objects from the Invisible AGN Catalogue. The complete catalogue is available online. Parenthetical values indicate 1$\sigma$ uncertainties.}

\label{catalogue}
\end{deluxetable}
\end{landscape}

Table \ref{catalogue} shows the first 10 objects from the Invisible AGN
Catalogue. The entire catalogue contains 46,258 objects with the same
columns and is available online and at http://vizier.u-strasbg.fr/. Each
catalogue entry contains information about the object's FIRST position and
flux density, its WISE position and magnitudes, and the separation between
the two positions. More information about each object's WISE and FIRST
detections can be found in the published AllWISE and FIRST catalogues.

From the FIRST catalogue, we list the object's FIRST ID, its position, its
position uncertainty, its
integrated 1.4 GHz flux density, and the local noise estimate at the object
position (Sky RMS). The SNR of the detection can be calculated from the sky
RMS and the peak 1.4 GHz flux density:
\begin{equation}
\mbox{SNR} = \frac{F_{\mbox{\mbox{peak}}} - 0.25}{\mbox{RMS}}.
\end{equation}
During the creation of the FIRST catalogue, 0.25 mJy is added to the peak
flux density to correct for the ``CLEAN bias'' effect ($\sim 0.25$ mJy of
flux density is lost during image CLEANing; \citealt{Beckeretal1994}). To
find the true SNR, this 0.25 mJy must be removed from the reported peak
flux density.  

The position uncertainties in FIRST are calculated from the uncertainties (90 per cent confidence)
in the source position ellipse:
\begin{equation}
\sigma_{\mbox{Maj}} = \mbox{Maj} \left ( \frac{1}{\mbox{SNR}} + \frac{1}{20} \right )
\end{equation}
\begin{equation}
\sigma_{\mbox{Min}} = \mbox{Min} \left ( \frac{1}{\mbox{SNR}} + \frac{1}{20} \right )
\end{equation}
where $\mbox{Maj}$ and $\mbox{Min}$ are the major and minor axes derived from the
elliptical Gaussian model for the source, before deconvolution. The $\frac{1}{20}$ addition to the uncertainties represents the estimated upper limit of systematic errors \citep{Beckeretal1994}. The position uncertainties of FIRST and their covariance are calculated using:
\begin{equation}\label{sigma_a}
\sigma_\alpha^2 = \sigma_{\mbox{Maj}}^2 \sin^2{\mbox{PA}} + \sigma_{\mbox{Min}}^2 \cos^2{\mbox{PA}}
\end{equation}
\begin{equation}
\sigma_\delta^2 = \sigma_{\mbox{Maj}}^2 \cos^2{\mbox{PA}} + \sigma_{\mbox{Min}}^2 \sin^2{\mbox{PA}}
\end{equation}
\begin{equation}\label{covar}
\sigma_{\alpha \delta}^2 = (\sigma_{\mbox{Maj}}^2 - \sigma_\alpha^2) \ \tan{\mbox{PA}}
\end{equation}
where PA is the position angle of the elliptical Gaussian model and $\sigma^2_{\alpha\delta}$ is the covariance of the FIRST coordinates \citep{Cutrietal2014}.

From the AllWISE catalogue, we list the object's AllWISE ID, its position, its position uncertainty, and its measured magnitudes, associated uncertainties, and SNRs in [3.4], [4.6], [12], and [22]. If SNR $<2$, the reported magnitude is an upper limit and no uncertainty is calculated \citep{Wrightetal2010}. AllWISE provides $\sigma_{\mbox{Maj}}$ and $\sigma_{\mbox{Min}}$ for each object, which we then use to calculate the position uncertainties and covariance using Equations \ref{sigma_a} -- \ref{covar}. Objects that are extended in AllWISE are marked with an
  asterisk after their AllWISE ID.  The reported magnitudes for these
  objects may under-represent the total source flux density and we
  recommend individual examination to ensure accurate flux measurements.

For an object with FIRST and AllWISE
positions of  $(\alpha_1,\delta_1)$ and $(\alpha_2,\delta_2)$, respectively, the angular separation between the positions from the two different observations is 
\begin{equation}\label{theta}
\theta = \arccos \left [ \sin \delta_1 \sin \delta_2 + \cos \delta_1 \cos \delta_2 \cos (\alpha_1 - \alpha_2 ) \right ]
\end{equation}
with associated uncertainty 
\begin{equation}\label{sigmatheta}
\begin{aligned}
\sigma_\theta^2 = \left (\frac{\partial \theta}{\partial \delta_1} \sigma_{\delta_1}
\right )^2 + \left ( \frac{\partial \theta}{\partial \alpha_1} \sigma_{\alpha_1}
\right )^2+ 2 \ \frac{\partial \theta}{\partial \delta_1} \frac{\partial
 \theta}{\partial \alpha_1} \sigma_{\alpha_1 \delta_1}^2 + \\ \left
(\frac{\partial \theta}{\partial \delta_2} \sigma_{\delta_2} \right )^2 + \left
( \frac{\partial \theta}{\partial \alpha_2} \sigma_{\alpha_2} \right )^2+ 2 \ \frac{\partial \theta}{\partial \delta_2} \frac{\partial \theta}{\partial \alpha_2} \sigma_{\alpha_2 \delta_2}^2
\end{aligned}
\end{equation} 
where $\sigma_{\alpha_1 \delta_1}^2$ and $\sigma_{\alpha_2 \delta_2}^2$ are the covariances of the coordinates (we assume no covariance between the FIRST and AllWISE positions). Many FIRST sources have a significant correlation between uncertainties in RA and Dec because the beam is often not orthogonal to the equatorial coordinate system. The partial derivatives in Equation \ref{sigmatheta} are given by
\begin{equation}
\frac{\partial \theta}{\partial \delta_1} = \frac{-1}{\sqrt{1-x^2}} \left [ \cos \delta_1 \sin \delta_2 - \sin \delta_1 \cos \delta_2 \cos (\alpha_1 - \alpha_2 )  \right ],
\end{equation}
\begin{equation}
\frac{\partial \theta}{\partial \delta_2} = \frac{-1}{\sqrt{1-x^2}} \left [ \sin \delta_1 \cos \delta_2 - \cos \delta_1 \sin \delta_2 \cos (\alpha_1 - \alpha_2 )  \right ],
\end{equation}
\begin{equation}
\frac{\partial \theta}{\partial \alpha_1} = \frac{1}{\sqrt{1-x^2}} \left [ \cos \delta_1 \cos \delta_2 \sin (\alpha_1 - \alpha_2 )  \right ],
\end{equation} and
\begin{equation}
\frac{\partial \theta}{\partial \alpha_2} =  \frac{-1}{\sqrt{1-x^2}} \left [ \cos \delta_1 \cos \delta_2 \sin (\alpha_1 - \alpha_2 )  \right ]
\end{equation}
where
\begin{equation}
x=\cos{\theta}.
\end{equation}


\bsp	
\label{lastpage}
\end{document}